# NOISE-TOLERANT STREAMING REAL-TIME DATA ACQUISITION FABRIC FOR PULSED ACCELERATORS

E. J. Siskind, NYCB Real-Time Computing, Inc., Lattingtown, NY 11560-1025, USA

## Abstract

A noise-tolerant data communications fabric has been developed to meet the real-time data acquisition and control requirements of fast feedback loops, machine protection systems, pulse-to-pulse sequencing, and machine-experiment communications at next-generation pulsed accelerators such as the Next Linear Collider ("NLC"). The fabric is constructed from "platform" or "system-on-a-chip" field programmable gate arrays ("FPGAs") containing embedded processors, block memory, and multi-gigabit serial transceivers interconnected via an array of point-to-point fiber-optic physical links for standard networks such as gigabit Ethernet. The FPGA-based link hardware segments messages of varying priorities into a continuous sequence of fixed-length data cells, interrupting the cell stream of lower priority messages with those comprising higher priority traffic. A high level of noise-tolerance is provided by dedicating almost half the contents of each cell to Reed-Solomon forward error correcting code ("ECC") data. Further reliability is achieved by hardware implementation of algorithms for cell-by-cell receipt acknowledgment, receipt timeout and retransmission after a tunable round-trip propagation delay, and separate backpressure flow control for each priority. Platform FPGAs containing a connection to this fabric can be used to implement the digital back ends of device controllers for RF stations, beam position monitors, machine protection input sensors, and fast-response power supplies for corrector magnets used in feedback loops. Additional FPGAs provide sector-by-sector data concentration as well as backplane interfaces for commercial off-the-shelf ("COTS") server PCs that can be located in the accelerator central campus. The resultant fabric is capable of streaming all pulsed waveforms present in front-end device controllers to the central campus COTS computing/networking installation in real time, while extending TCP/IP connectivity at non-real-time priority to all pulsed device controllers connected to the fabric. The bandwidth provided by concentrating the data from one currently available fiber per device controller into 1-2 10-gigabit/second links per accelerator sector is adequate to meet all data acquisition and control requirements.

## 1 INTRODUCTION

Experience at the SLAC Linear Collider ("SLC") strongly highlights the wisdom of integrating support for real-time data communications into the architecture of the control system for a pulsed accelerator *ab initio*. The failure to provide adequate real-time support in the networking infrastructure has led to the addition of separate *ad hoc* networks or communications paths for each of fast feedback, machine protection, pulse-to-pulse sequencing, and machine-experiment communication. This approach has deleterious effects on control system reliability, availability, and maintainability; it also increases the costs of hardware and software development, deployment, and maintenance.

The SLC experience also highlighted the problems with radio-frequency interference ("RFI") generated by high-power pulsed energy generation and distribution in systems such as klystron modulators and kickers. The effects of RFI are manifested as noise pickup in sensitive high bandwidth electronic systems in the general vicinity of the pulsed energy systems. These problems are especially severe when signals travel from one ground plane to another, and are most noticeable when connecting one circuit board to another via a parallel backplane or a serial communications link. However, they may be present to a lesser extent when moving from one substrate to another, i.e. when routing signals from one integrated circuit to another via printed circuit board traces. The last several decades of research into information theory and communications suggests that noise can never be completely eliminated from communications systems. This implies that when the effects of noise are intolerable, at some point it becomes advantageous to invest incremental effort in achieving noise tolerance rather than additional noise reduction. Since parallel backplanes and similar parallel connections are susceptible to protocol violations as well as addressing errors and data corruption in the presence of noise, this in turn suggests building a data communications architecture for a pulsed accelerator around noise-tolerant serial links.

Fortunately, the tools for achieving acceptable real-time latency and noise tolerance in serial links have been around for several years, and are increasingly accessible to the run-of-the-mill engineer. In particular, the tools for allocating guaranteed portions of the bandwidth available on a given link to individual circuits on a prioritized basis are present in recently developed standards for circuit-switched networks such as ATM. These generally consist of dividing the data packets sent on each circuit into smaller cells, and then transporting the cells on a prioritized basis. While the tools for achieving noise tolerance via Reed-Solomon ECC precluded their implementation on even the largest FPGAs available in the mid-1990s, by 2000 the required logic could be implemented within an FPGA with ease. In addition, on that time scale it became possible to generate the intellectual property necessary to implement a Reed-Solomon encoder or decoder with customized parameters in a few seconds via a graphical interface on a PC. In mid-2002, platform FPGAs integrating programmable "glue" logic with embedded RISC processors, block memory, and multi-gigabit serial transceivers became available. These appear to be the ideal platform on which to implement the digital back ends of device controllers in any accelerator. The volume of FPGA fabric available within these parts has now become so large that the incremental portion that is required to achieve noise tolerance and well-defined real-time latency on their communications links is marginal, and usually fits in the unused portion of any given part.

The remainder of this paper describes a particular *ad hoc* data communications fabric developed for pulsed accelerators and a proposed real-time data streaming architecture constructed on this fabric. The hardware of each communications link in the fabric is circuit switched among just three circuits. The highest priority circuit transports fixed-length messages to/from processor-accessible registers at the link ends. This "trigger pattern" circuit is dedicated to data controlling pulse-to-pulse sequencing of accelerator operation. The remaining two circuits forward variable length messages to/from independently specified locations in processor-accessible memory at each end of the link. The lower priority message circuit carries traffic for a beam-asynchronous packet-switched network. In the long term, the PPP protocol will be implemented on this circuit, providing TCP/IP connectivity to device controllers connected to the fabric. This will permit new or existing TCP/IP-based control system software to communicate with the device controllers to effect beam-asynchronous control functions and data readout. The higher priority message circuit carries the traffic for a beam-synchronous network. Packet switching software will be used to multiplex the individual real-time services onto this common circuit. The requirements of those services are closely aligned with the capabilities provided by the UDP protocol; there is no *a priori* reason why this circuit can't be used to implement a separate beam-synchronous UDP network that is distinct from and higher priority than the TCP/IP network on the lower priority message circuit. There is no reason to implement the full TCP/IP protocol on the beam-synchronous circuit because the underlying fabric hardware provides flow control, hardware acknowledgment, and acknowledgment timeout and retransmission services.

## 2 LINK CHARACTERISTICS

The characteristics of the cells into which link traffic is divided must be chosen to take into account both the characteristics of anticipated RFI sources and the bandwidth of the physical link employed. In general, RFI is only generated on the rising and falling edges of pulses. Therefore, any cell as transported on the physical link must be either longer than the full width of the expected pulses and tolerant of noise generated on both pulse edges, or else shorter than the pulse width and tolerant of noise on one entire edge. The latter choice is preferable because shorter cells lead to lower real-time latency.

In general, although the design of a Reed-Solomon decoder is pipelined, it cannot begin accepting a new code block until after a processing delay has elapsed. This delay is measured from the beginning of the data input process for the current code block, and features a leading term that is quadratic in the number of ECC bytes contained in the block. This behavior results from the fact that the processing delay is associated with calculating (and in some cases solving) the coefficient matrices for the systems of equations that must be solved to decode the block. The dimensions of these matrices are equal to the number of ECC bytes. Ideal hardware utilization is achieved when the length of the code block is chosen to be equal to the processing delay, so that the input of the next code block can commence immediately after the input of the previous block has been completed. Note that the processing delay is distinct from, and shorter than, the decoder's output latency. This is the time required to calculate the matrices, solve the equations, and apply any necessary corrections to the bytes in the code block that was originally input.

In the initial implementation of the link, the cell characteristics were selected to anticipate the noise environment in the SLAC LINAC with data transport on gigabit Ethernet physical link hardware. This choice resulted in a cell that was 4.864 microseconds or 608 bytes long, and composed of 32 interleaved Reed-Solomon (19,11) ECC blocks. Each of the 32 code blocks consists of 10 bytes of data payload, 1 byte of hardware header, and 8 bytes of Reed-Solomon ECC data. Each code block is tolerant of the loss of any four arbitrarily positioned bytes. A noise burst that destroys the contents of 127 consecutive bytes of data within one cell, positioned at an arbitrary phase with respect to a byte boundary, will corrupt no more than 4 bytes in any one of the 32 ECC blocks. The link is thus tolerant of a noise burst in each cell that is 127 bytes or 1.016 microseconds long. In summary, the selected characteristics result in a link that has a real-time latency of less than 5 microseconds, is tolerant of a microsecond noise burst every 5 microseconds, and has a limiting bandwidth of 10/19 of the underlying physical link bandwidth, or 65.79 megabytes/second. This bandwidth is achieved if all messages have a length that is an integral multiple of the 320-byte data payload of a single cell.

The link hardware utilizes a three-stage pipeline at both transmitting and receiving ends, with double buffers for cell storage between successive stages. At the sending end, cells are constructed in the first stage, have their ECC bytes added in the second stage, and have their 8B/10B encoding and serialization functions performed in the third stage. The third stage inserts a long sequence of known two-byte 8B/10B ordered sets, each of which consists of a K28.5+ "comma" character followed by a D21.4- character, to initialize the link. This permits the deserializer at the receiving end to locate the first bit in each encoded byte, as well the first byte in each two-byte pair. Once the link is initialized, the feature in the deserializer that detects comma characters and re-synchs the first bit in a byte is disabled. As a result, the link is no longer sensitive to noise that mimics comma characters disrupting the data flow. The third stage of the transmit pipeline can also insert a one-byte delay in the outgoing data stream so that the first byte of the first outgoing cell always follows the last D21.4- character rather than the last comma. Once this action is performed, the process of sending a continuous stream of cells ensures that there are no further "start-of-message" sequences conveyed outside of the region protected by Reed-Solomon ECC that can be corrupted by noise. The pipeline at the receiving end performs deserialization and 8B/10B decoding in the first stage, Reed-Solomon decoding in the second stage, and cell disassembly and header processing in the third stage.

All stages at the transmitting end of the link are clocked by a local 125 MHz source. A single Reed-Solomon encoder clocked at 125 MHz is employed to encode all 32 of the code blocks. Cell data are written into the double buffers between the first and second transmit stages in linear order but are read out in interleaved order. Similarly, encoded cells are written into the buffers between the second and third transmit stages in interleaved order but are read out in linear order. A clock that is recovered by the deserializer from the incoming bitstream clocks the first two stages at the receiving end at 62.5 MHz. This 62.5 MHz clock is phase-locked with the local 125 MHz clock at the transmitting end. The third stage of the receiving pipeline is clocked by the local 125 MHz source at the receiving end. The transition between clock domains occurs in the double buffers between the second and third stages at this end. Necessary elasticity is provided by the fact that the duty cycle in the incoming data stream is far less than 100% once the ECC bytes are removed.

The current implementation employs eight Reed-Solomon decoders at the receiving end with each decoder processing four of the 32 code blocks in each cell. Each decoder accepts one new input byte every four ticks of the 62.5 MHz clock; the combined set of 8 decoders, each processing with an effective clock rate of 15.625 MHz, has the same throughput as the single 125 MHz encoder at the transmitting end. This choice is simply a reflection of the current capabilities of existing FPGA families and is not an immutable link feature, even if the same cell characteristics were selected.

## 3 LINK HEADER FORMAT

The combination of the single bytes of header data included in each of the 32 interleaved code blocks forms an 8-longword header for each cell. The first longword contains descriptive control information about the cell, while the second longword provides the write data for the implementation of write access to the link initialization register at the far end of the link. This allows the transmitter at one end of the link to initialize the transmitter at the other end of the link, as well as any embedded intelligence at the far end. The 64 bits in the next two longwords implement a full four-way request-grant handshake implementing backpressure flow control for 16 receive buffers on each of the beam-synchronous and beam-asynchronous circuits. No flow control *per se* is provided for the registers that receive incoming trigger patterns. This function is implicitly provided because the request to forward a trigger pattern is not cleared until the far end of the link acknowledges receipt of the cell carrying these data (or its retransmitted equivalent). A similar interlocking function is provided for writing the far end's link initialization register.

The remaining four header longwords carry the current contents of eight 16-bit counters. The first of these carries the sequence number of the transmitted cell, while the second carries the sequence number of the next expected incoming cell. This is one greater than the sequence number of the last cell that was received in proper sequence order with no uncorrectable Reed-Solomon decoder errors. The remaining six counters record various errors including 8B/10B decoder errors, correctable and uncorrectable Reed-Solomon decoder errors, cells received with unexpected sequence numbers, timeout events, and cells received with unexpected values for the next expected incoming sequence number. This last error count is not necessarily indicative of any problems, since the incoming link can process more or less than one incoming cell during the time between departure of successive outgoing cells because of the difference between local 125 MHz clock frequencies at the two ends of the link.

The cell control information longword at the beginning of each cell's header contains a 16-bit control field and a 16-bit data offset field. The low order four bits of the control field contain the destination receive buffer number for the message of which the cell forms a portion. This field is only valid when the cell forms a part of the data stream on the beam-synchronous or beam-asynchronous circuit. The next two bits identify the circuit, specifying one of these two circuits, the trigger pattern circuit, or a no-op cell containing only header and ECC data but no valid data payload. The next bit identifies the first cell in a new message buffer on one of the two message circuits, and is used to force the DMA engine at the receiving end to reload its memory address register with the initial address of the message buffer for the appropriate circuit specified by the low four bits. The next bit identifies the last cell in a message buffer, and is used at the receiving end to update the flow control handshake lines for the current message buffers to reflect the new "unavailable" state of the buffer just filled. It is also used to interrupt the local processor at this end if so requested. When receipt of this cell is acknowledged, the transmitting end uses this bit to clear the request to transmit the message and interrupt the local processor if so requested. The next bit is the write enable bit for the far end link initialization register. When receipt of this cell is acknowledged, the transmitting end uses this bit to clear the request to write the far end link initialization register. The remaining seven bits of the control field are unused. The data offset field contains the cumulative amount of data in the current message buffer that has been transmitted, including the contents of the data payload portion of the current cell. The value contained in the last cell in a message is the total message length, and is used to load the received byte count register for the appropriate message buffer. This field is expressed in units of 64-byte chunks, i.e. five units for the 320-byte data payload of each cell.

## 4 LINK TIMEOUT FUNCTIONALITY

As each cell is transmitted, its cell sequence number is entered into a 48-bit-wide FIFO along with the contents of its control field and a value of its data offset field excluding the contents of the current cell. The contents of the cell sequence number field at the FIFO output are continually compared to the next expected incoming sequence number field contained in the header of cells received from the far end of the link. If the next expected sequence number from the far end exceeds the sequence number at the FIFO output (but not by more than the timeout value) then the cell described by the FIFO output must have been properly received at the far end. The FIFO entry may be purged and any message, pattern, or far end initialization register write data transmission requests completed by this cell may be marked as having been accomplished.

If the cell sequence number field currently being entered into the FIFO exceeds the value of this field at the FIFO output by more than the timeout value, then an outgoing cell must have been lost. In this case a timeout and cell-retransmission event is initiated. The contents of the transmitted cell sequence number counter are reloaded from the cell sequence number field at the FIFO output, and the contents of the other 32 bits of the each word in the FIFO are transferred to a second FIFO. The cells described by the contents of this second FIFO are then retransmitted, while re-entering the descriptions of these cells into the first FIFO. The contents of the control longword, transmitted cell sequence number counter, and data payload of each retransmitted cell are identical to their original values. However, the contents of all other header fields reflect their updated values. A new timeout and cell-retransmission event may be declared if no cell descriptions have been purged from the first FIFO when the second FIFO has been emptied. Hopefully this will not be the case unless the link has been disconnected or power removed from the far end.

At the SLAC LINAC, the typical value for the timeout value is 20 cell lengths or 10 cell lengths for propagation in each direction, corresponding to a timeout delay of ~100 microseconds. The 10 cell lengths of allowable unidirectional delay consist of ~6 cells for passage through transmit and receive pipelines plus ~4 cells for the actual time-of-flight, corresponding to a link length of ~20 microseconds or ~6 kilometers. Note that the last stage of the transmit pipeline overlaps the first stage of the receive pipeline in a zero-length link, but that the second receive stage has a delay longer than the cell length because of the Reed-Solomon decoder's output latency.

## 5 LINK IMPLEMENTATIONS

The first link implementation was realized in a Xilinx XCV1000E Virtex-E series FPGA in a 680-pin package. The link's logic consumes 5-6k logic slices plus a small number of block memories (mostly for the double buffers between cell-processing pipeline stages) on this platform. The eight Reed-Solomon decoders consume the majority (~4k slices) of this logic. With the aid of external static RAM, an Analog Devices ADSP-21060 SHARC processor, and external Agilent HDMP-1636A serializer/deserializer and HFCT-53D5 single-mode fiber-optic transceiver for gigabit Ethernet, this provided an intelligent interface between a COTS PC and the link. This circuitry was packaged on a 64-bit/66-MHz PCI board, interfacing to the PCI via a Xilinx PCI core compiled into the FPGA. The PCI master interface is capable of full-speed transfers between PCI-accessible memory and on-board memory at 533 megabytes/second. As perceived from the SHARC, there are two logically independent PCI masters, one for beam-synchronous and one for beam-asynchronous I/O. Operation of the synchronous master interrupts the asynchronous master with a latency of no more than 3 PCI data words. This board became operational early in 2002.

The link is currently being re-implemented on a Xilinx XC2VP50 Virtex-II Pro platform FPGA with embedded PowerPC processors and multi-gigabit serial transceivers. Utilization of this platform eliminates the requirement for the external SHARC processor and Agilent serializer/deserializer component. On the Virtex-II Pro platform, the implementation of the link hardware has become a simple piece of intellectual property that forms a DMA device on the PowerPC's main memory, with control functions implemented by registers on the PowerPC's Device Control Register bus. The first use of a Virtex-II Pro component containing the link interface is in an intelligent triggered remote I/O controller for the legacy CAMAC and Bitbus field busses in the SLAC LINAC and PEP-II storage rings. This device will allow a COTS PC sited in the computer room at the SLAC Main Control Center to access the device controllers connected to the existing field bus plant in any one of the ~80 sectors currently served by the SLAC control system. It is hoped that deployment of hardware implementing this architecture will commence in the third quarter of 2003, replacing the aging Multibus-I microcomputer systems now deployed in each sector.

## 6 FABRIC CONSTRUCTION

A real-time data acquisition fabric provides a means of connecting one or more COTS computers to an assortment of pulsed device controllers in an accelerator sector. Construction of a fabric requires the design of one more class of device in addition to the existing interfaces between device controllers or COTS PC backplanes and noise-tolerant streaming real-time data links. This is a switching node providing a connection among three or more such links, possibly with differing bandwidths. Fortunately, FPGAs scheduled to reach the market by the end of 2002 seem to provide an ideal platform on which to construct not only device controllers but also fabric switching nodes and interfaces between noise-tolerant streaming real-time links and both current parallel and future serial backplanes for COTS PCs.

This situation arises because already announced members of the Virtex-II Pro family embed multiple PowerPCs and multiple high bandwidth serial transceivers in each part's FPGA fabric. Each PowerPC is capable of clocking at advertised rates of at least 300 MHz. There are four such processors embedded in the largest components in the Virtex-II Pro family. Similarly, each serial transceiver is capable of 8B/10B encoding a data stream of 2.5 gigabits/second onto a 3.125 gigabit/second link. Furthermore, up to four such links can be bonded together to form a single coherent transport mechanism for 10 gigabit Ethernet or similar bandwidth protocols. The largest Virtex-II Pro parts available by the end of the year will contain 24 such serial transceivers, along with the 4 PowerPCs, 556 18-kbit block memories, and more than 55k logic slices. These components provide up to 1,200 parallel user I/O pads in addition to their serial links, and are packaged in flip-chip ball-grid-arrays containing up to 1,704 pins. Future families of platform FPGAs can only be larger and faster.

The 5-6k slices required in order to support the single noise-tolerant streaming real-time link on a device controller or COTS PC interface will require an ever-smaller percentage of the FPGA fabric provided in ever-larger families of FPGA components. In addition, the volume of logic required to implement a fixed bandwidth link will fall substantially as faster logic speeds permit the number of Reed-Solomon decoders required per link to decrease. New and existing families of platform FPGAs should continue to provide an ideal means of interfacing the data acquisition fabric to COTS PCs even as the I/O busses employed by these systems migrate from parallel to serial protocols. This is true because the multi-gigabit serial transceivers embedded in these chips will support the standard serial backplane protocols employed in future PCs. Already Virtex-II Pro components have implemented some of the first interfaces to the emerging PCI Express standard, as well as supporting Fibre Channel and Infiniband I/O interconnects.

The presence of 24 multi-gigabit serial transceivers in larger Virtex-II Pro ports would seem to provide an ideal platform for implementing a switching node in the data acquisition fabric. Such a node could merge the data flowing from 20 slower links onto one faster link constructed by bonding four transceivers together. Alternatively, data flow from 16 slower links could be merged onto two faster links. This would provide a tool for concentrating data from all pulsed devices in an accelerator sector onto 1-2 links connected to a COTS PC in the central campus. An alternative purpose might be identical except that it is viewed from the other end, i.e. to distribute data from one faster link into multiple slower links. Unfortunately, the current ratio of logic slices to serial transceivers is a bit too small to permit the implementation of noise tolerance on more than ~8 links connected to such a switching node. As logic densities increase, it will no doubt become possible to construct a single-chip implementation of a switching node to merge the data from all pulsed device controllers in an NLC sector onto a single noise-tolerant streaming real-time link.

## 7 DATA ACQUISITION FEATURES

In considering the data flow architecture in a pulsed accelerator, it is important to recognize that at least some portion of the pulse-to-pulse data available from pulsed device controllers in each accelerator sector must be acquired in real time for use in fast feedback loops, machine protection systems, or similar real-time applications. The problems encountered in reliably acquiring these data, calculating meaningful results from them, and utilizing these results to adjust machine parameters for the next pulse in an environment where a single incorrectly steered pulse may cause significant damage to the accelerator are considerable. These problems are distinct from, orthogonal to, and subject to solution by different techniques than the not-inconsiderable challenges of providing user-friendly access to the vast number of non-pulsed devices in a large accelerator plant. In fact, the problems of pulsed accelerator real-time data acquisition, processing, and dispersal are far more similar to those faced in data acquisition for experiments at pulsed accelerators than they are to the data access problems of control systems for non-pulsed machines. In fact, it is most useful to consider the pulsed devices in an accelerator as comprising just another detector for an experiment running at that accelerator. This detector has the added feature that its event rate is absolutely periodic rather than stochastically distributed, thus eliminating the need for some rate-smoothing buffering. However, it has the added constraint that the data acquisition pipeline cannot be arbitrarily deep, since data acquired from each event are required to influence detector parameters for the next event.

It is also useful to note that the data flows in any real-time application are topologically equivalent to those in a single feedback loop or a subset of such a loop. By this it is meant that they comprise one or more of data acquisition, computation, and data dispersal, as well as the relaxation time of corrector devices, all within a single accelerator pulse period. Thus, the topological features of data flow in a machine protection system are identical to those in a global feedback loop. Similarly, accelerator-experiment communication of pulse-to-pulse data involves data acquisition and data dispersal, if not actual computation.

A considerable amount of time was invested in determining what fraction of the pulse-to-pulse data available in each pulsed device controller should be streamed via the real-time circuit provided by the data acquisition fabric to the COTS PC associated with its accelerator sector. In particular, it was initially unclear what should be done with pulse-to-pulse data for which no real-time demand was currently extant. After discussions with the group responsible for construction of the data acquisition software for SLD, the last large experiment at a pulsed accelerator, it became clear that the simplest software architecture is achieved when all pulsed data available to device controllers are unconditionally pushed towards the central campus COTS computers via the real-time circuit. This statement is obviously only valid if sufficient bandwidth exists for this purpose in the data acquisition fabric, but holds true regardless of whether a demand for any individual data item currently exists within the COTS environment. This does not preclude the possibility that some non-real-time software application may wish to fetch data directly from a pulsed device controller via TCP/IP transported on the lower priority circuit in the data acquisition fabric. However, an application should be constructed in this fashion only if doing so simplifies its construction and does present unacceptable bandwidth loads to the fabric.

The rationale behind this assertion lies in the fact that the software at the receiving end of a data stream must always be cognizant of all demands that currently exist for each data item in the stream, and capable of distributing each data item according to its current demands. This is true even if there are no current demands, i.e. the only sink for the data item is the bit bucket. Thus, the complexity of this section of the software does not change regardless of whether data are streamed towards the central campus only when required, or under all circumstances. On the other hand, the complexity of the software at the sending end, i.e. in the pulsed device controllers, is significantly impacted by the choice of conditions under which data items are forwarded via the data acquisition fabric. In particular, it is far simpler to send data items as soon as they become available rather than wait for a request for the items. In other words, it is easier to push out data as soon as they become available rather than wait for someone to pull the data out of the controller. Similarly, it is far easier to construct software that unconditionally pushes out the data than it is to provide the mechanisms necessary to dynamically construct the data structures required to specify that a given data item should be pushed onto the fabric, and then link these structures into the appropriate database. Conditional specification of the data items to be forwarded onto the fabric also requires dynamic unlinking of control structures and their return to the heap once a particular data item is no longer required at the destination. In effect, streaming all pulse-to-pulse data in real-time costs nothing as long as bandwidth is available, and leads to simpler software in the data sources.

## 8 STREAMING DATA BANDWIDTHS

In an example of a SLAC version of an NLC of some recent design generation, the RF plant of each accelerator sector contains 9 RF stations. Each station employs 8 klystrons and feeds the RF structures supported on 8 girders via a common delay-line distribution system. A rational estimate of the data volume required to contain digitized versions of all pulsed waveforms available to the device controller for a single station is 256 kilobytes per pulse. This estimate consists of 4 sampled waveforms with 4 kilobytes per waveform associated with each klystron, plus 8 sampled waveforms with 2 kilobytes per waveform associated with each girder. Girder waveforms contain fewer samples because pulse widths are shorter at girders than at klystrons. Note that these numbers are not authoritative, but are simply advanced to establish plausibility. For comparison, the bandwidth of the first implementation of the noise-tolerant link on gigabit Ethernet physical hardware (with parameters tuned to the SLAC LINAC) is 535 kilobytes/pulse at a 120 Hz pulse rate. Thus, connecting each device controller for an RF station to the data acquisition fabric via a single gigabit Ethernet fiber, or certainly via the 2.5-gigabit/second serial transceivers that are already extant in Virtex-II Pro FPGAs, should prove adequate.

The volume of data associated with pulsed waveforms generated by beam position monitors is somewhat less than that associated with RF stations. The maximum rate of instrumentation is one X-Y monitor per girder. If one assumes 4 kilobytes of data per axis and combines the data flow from 18 girders into a single fiber, then the real-time data flow in that fiber is 144 kilobytes/pulse. The combined data volume from 9 RF stations and 4 beam position monitor controllers is thus estimated at 2,880 kilobytes per pulse. The remaining pulse-to-pulse data consist of single analog-to-digital conversions and single digital status bits associated with machine protection inputs heading inwards towards the central campus, plus a similar volume of single parameters produced by feedback loops heading outwards from the COTS plant. This suggests that one or at most two noise-tolerant links per accelerator sector implemented on full-duplex media at 10 gigabits/second should be sufficient to stream digitized versions of all pulsed waveforms present in the sector to the central campus in real time.

One can similarly estimate the volume of pulse-to-pulse data generated by the entire accelerator as being of the order of magnitude of 100-400 megabytes/pulse (50-100 accelerator sectors at 2-4 megabytes/pulse/sector) at a pulse rate of 1 gigapulse/accelerator running year. This is 100-400 petabytes per accelerator running year. Entering this volume of data into a formal database is beyond the capacity of any rationally priced system that can be constructed with existing technology. However, it is certainly not beyond reach by the time that accelerator commissioning, detector shakedown, or serious experimental data taking are commenced.

By the time that accelerator commissioning occurs, it should easily be possible to equip a COTS PC receiving streaming data from one accelerator sector with at least 128 gigabytes of low-cost main memory. At a 120 Hz pulse rate and 2-4 megabytes/pulse/sector, this is something like 32k machine pulses or close to five minutes of pulse-to-pulse data. This should prove useful when diagnosing the causes of aberrant single-pulse or transient behavior in the accelerator. It is surely less expensive to incrementally increase the depth of this silo by upgrading the COTS PCs in the central campus every 2-3 years than it is to retrofit all of the device controllers distributed throughout the accelerator with higher capacity local memory. Similarly, the lead-time for engineering, constructing, and installing pulsed device controllers is far longer than that for procuring the initial plant of COTS machines. Thus, even this initial COTS plant will use newer technology than the device controllers. This argues strongly for implementing such a silo in COTS hardware in a central location rather than in special-purpose distributed hardware such as accelerator device controllers.

Current technology is (perhaps optimistically) capable of making database additions at 1 petabyte/running year. For example, the BaBar collaboration recently announced that its database had reached 0.5 petabytes, with the majority of these data added relatively recently as accelerator luminosity has increased. Presumably the database manager being developed for experiments on the Large Hadron Collider at CERN will be capable of something like an order of magnitude greater performance, although no attempt has been made to verify this assertion. Certainly, if Moore's law continues to apply with a doubling time of 1.4 to 2.0 years then in 14 years capacities should increase by 2-3 orders of magnitude, reaching if not exceeding the desired benchmark. This adds to the argument that, if it is relatively easy to design the data flow architecture to be capable of streaming all pulse-to-pulse waveforms available in an accelerator to a COTS computing facility in a central campus by utilizing techniques and technologies already available, then one should do so. This conclusion applies even if one must route most of these data into the bit bucket for the immediate future. In the long run, one's capacity for utilizing and storing these data can only grow, making the choice to acquire all pulsed data in real time at minimal cost from the outset appear far-sighted.

## 9 ACKNOWLEDGEMENTS

Discussions on real-time data flow architectures in experiments at pulsed accelerators with J. J. Russell, M. Huffer, and J. Bogart are gratefully acknowledged. This research was funded by the U.S. Department of Energy ("DOE") under SBIR grant DE-FG02-98ER82628. Additional DOE funding was provided by SLAC's operating contract and its subcontracts.